\documentstyle[twoside,fleqn,npb,epsfig]{article}
\def\beq{\begin{equation}}
\def\eeq{\end{equation}}

\def\beqn{\begin{eqnarray}}
\def\eeqn{\end{eqnarray}}

\def\np#1#2#3  {{\it Nucl. Phys. }{\bf #1} (19#3) #2} 
\def\nc#1#2#3  {{\it Nuovo. Cim. }{\bf #1} (19#3) #2} 
\def\pl#1#2#3  {{\it Phys. Lett. }{\bf #1} (19#3) #2} 
\def\pr#1#2#3  {{\it Phys. Rev. }{\bf #1} (19#3) #2} 
\def\prl#1#2#3  {{\it Phys. Rev. Lett. }{\bf #1} (19#3) #2} 
\def\prep#1#2#3 {{\it Phys. Rep. }{\bf #1} (19#3) #2} 
\def\zp#1#2#3  {{\it Z. Phys. }{\bf #1} (19#3) #2} 
\def\rmp#1#2#3  {{\it Rev. Mod. Phys. }{\bf #1} (19#3) #2} 
\def\hepph  #1 {{\tt hep-ph/#1}}

\newcommand\sss{\scriptscriptstyle}

\newcommand{\AmS}{{\protect\the\textfont2
  A\kern-.1667em\lower.5ex\hbox{M}\kern-.125emS}}
\title{Jet Production with Polarized Beams at Next-to-Leading Order}

\author{Daniel de Florian
\address{Institute of Theoretical Physics, ETH\\ CH-8093 Z\"urich, Switzerland}%
        \thanks{Work partly supported by the EU Fourth Framework Programme `Training and Mobility of Researchers', Network `Quantum Chromodynamics and the Deep Structure of Elementary Particles', contract FMRX-CT98-0194 (DG 12 - MIHT).} }

\begin{document}

\begin{abstract}
Jet production cross-sections in polarized proton-proton and electron-proton collisions are studied to next-to-leading order accuracy. Phenomenological results are presented for RHIC and HERA kinematics.
\end{abstract}

\maketitle

The last decade has seen an important advance in our understanding of
polarized nucleon structure functions as a result of the analysis of 
deep inelastic scattering (DIS) data. Unfortunately, the 
use of DIS data alone does not allow an accurate 
determination of the polarized parton densities.
This is true in particular for the gluon, since this quantity
contributes to DIS in leading order (LO) only via the $Q^2$-dependence 
of the spin asymmetry ($A_1^N$). 

At variance with DIS, collider physics offers a relatively large 
number of  processes whose dependence upon the gluon density is
dominant already at LO. The study of these processes is therefore
crucial in order to measure this density in a direct way.
Among them, jet production is an obvious
candidate, because of the large rates.

In order to make reliable quantitative predictions for  high-energy processes,
it is crucial to determine the NLO QCD corrections
to the Born approximation. The key issue here is to check
the perturbative stability of the process considered.
Only if the corrections are under control can a process that shows 
good sensitivity to, say, $\Delta g$ at the lowest order be regarded 
as a genuine probe of the polarized gluon distribution and be reliably 
used to extract it from future data.
NLO QCD corrections are expected to be particularly important for the
case of jet-production, since it is only at NLO that the QCD structure 
of the jet starts to play a r\^{o}le in the theoretical description, providing 
for the first time the possibility to realistically match the experimental 
conditions imposed to define a jet.

The main purpose of this talk is, therefore, to study the perturbative stability and the phenomenological consequences of jet production 
cross-sections to NLO accuracy.
We have implemented 
 the NLO QCD corrections to jet production in polarized
proton-proton and electron-proton (in the photoproduction regime) collisions 
by extending the unpolarized MonteCarlo code constructed in
 ref.~\cite{Jets97} to the case of polarized beams.  As a result, we  
present a customized code, with which it is possible to calculate 
any infrared-safe quantity corresponding to either single- or di-jet 
production to NLO accuracy. For the theoretical details about the implementation, we refer the reader to  refs. \cite{Jets97,sust,jetpp,jetep}.


The
best way to analyze the effect of NLO corrections on the 
perturbative stability of an observable   is to study 
 the dependence of the full NLO  result
on the renormalization and factorization scales. Throughout we will set
the two scales equal, i.e. $\mu_R = \mu_F \equiv \mu$, and vary $\mu$ as a way to quantify the  theoretical uncertainty on the cross section. 
 In fig.~1a we show  
the next-to-leading and leading order $p_{\sss T}$-distributions for polarized  $pp$ collisions for the three different scales: $\mu_0, 2 \mu_0 $ and $\frac{1}{2} \mu_0 $.
 Figure~1b is the corresponding plot for the
unpolarized case. Clearly, the dependence on the scale is substantially
reduced when going to next-to-leading order. The situation in the
polarized case is indeed very similar to the unpolarized  one. We have observed the same reduction in the scale dependence for other single and double-differential observables. It is 
therefore sensible to use our code to investigate  a 
few phenomenological issues relevant to hadronic physics at RHIC.
\begin{figure}
\centerline{
   \epsfig{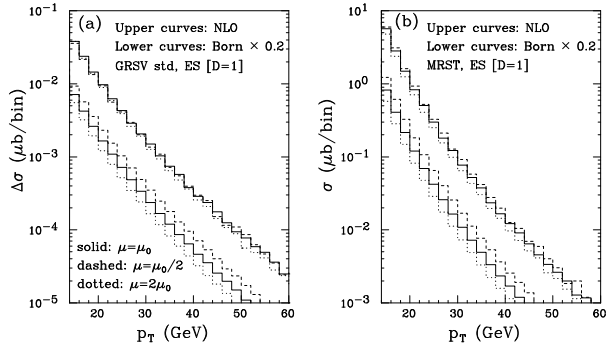} }
\vspace{-0.5cm}
\caption{ Scale dependence of the next-to-leading order and Born 
   $p_{\sss T}$-distributions in $pp$ collisions.}
\vspace{-0.4cm}
\end{figure}                                                              
 In fig.~2, 
the one-jet asymmetry
 is shown  as a function of $p_{\sss T}$. The
results  have been obtained by
choosing six different parametrizations of the polarized parton
densities \cite{polpar}.  Figure~2 clearly shows 
that the choice of the polarized parton densities induces an 
uncertainty on the theoretical results of more than two orders of 
magnitude. 
 This enormous spread is basically due to the fact
 that at this energy the jet cross section is
 dominated by $gg$- and $qg$-initiated parton processes.
Therefore, and since the minimum value of the asymmetry observable 
at RHIC is quite small,  the measurement of the polarized
jet cross section at RHIC will be useful in order to rule out
some of the polarized sets that are at present consistent with
the data.
\begin{figure}
\hspace{-.3cm}
\centerline{
   \epsfig{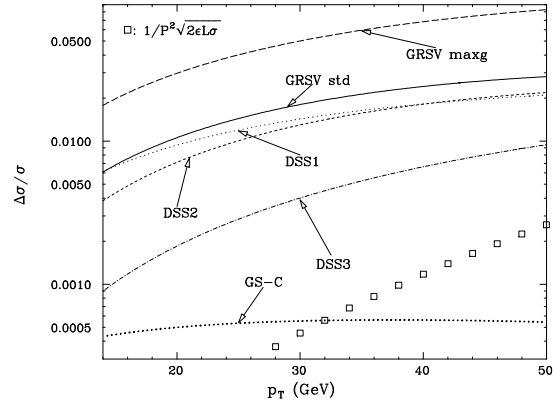} }
\vspace{-0.5cm}
\caption{ Single-inclusive jet asymmetry  in $pp$ collisions for various
polarized parton densities.}
\vspace{-0.4cm}
\end{figure}                                                              
The Born results differ from those presented in fig.~2
for a factor up to 20\% and the shape is
also different. Therefore, NLO corrections give non-trivial information
on the structure of the asymmetries


Moving to the case of electron-proton collisions, it is clear that in
 order to obtain large statistics one should go to the 
photoproduction regime. In this case the cross-sections can be approximated as a convolution of the photon-proton cross sections with the Weizs\"acker-Williams  flux. The photon-proton cross section is given by a sum of two terms, denoted as the point-like and hadronic components.
In order to compute the hadronic component one needs, besides the polarized parton distributions in the proton, also the polarized densities in the photon , which are completely unmeasured so far. To obtain a realistic estimate for
the theoretical uncertainties due to these densities, 
we use the two very different scenarios considered in
ref.~\cite{svgamma}.

We have studied the scale dependence of several observables at NLO and found that the variation of the scale induces a variation of the
cross section of the order of 10\% over most of the
$\eta$ range considered. The scale dependence is also strongly reduced
when going from Born  to NLO results.

We now turn to the problem of studying the dependence of our
results upon the available proton and photon polarized parton
densities.  In fig.~3 we present the results for the asymmetry
in terms of the  pseudorapidity of the single-inclusive jet 
at polarized HERA, obtained using GRSV STD as the polarized proton set and  both  polarized photon sets. The Born and NLO results are both shown. We can see that
in the large  $\eta$ region the difference induced by
the choice of the two photon sets is extremely large. On the
other hand, towards negative $\eta$ values this difference
tends to vanish. This is because in that region the point-like
component, which does not depend upon photon densities, is
the dominant one. We can also observe that in the positive $\eta$
region there is a very small difference between the NLO and
LO results, while for negative $\eta$'s the radiative corrections
are positive and reduce the asymmetry considerably. 
\begin{figure}
\centerline{
   \epsfig{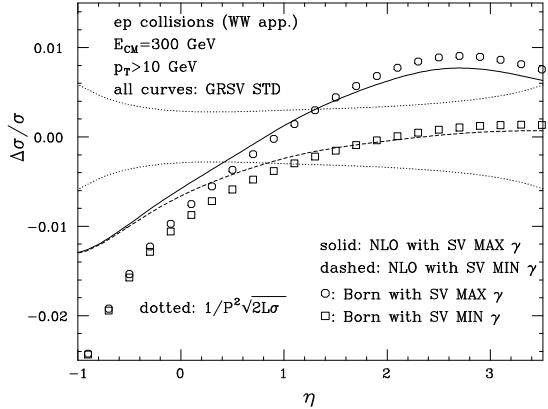} }
\vspace{-0.6cm}
\caption{
Asymmetries in single-inclusive jet production in electron--proton collisions. }
\vspace{-0.6cm}
\end{figure}                                                              
In fig.~4, we show the curves obtained
by fixing the polarized photon set to \mbox{SV MAX $\gamma$}, and
by considering the various polarized proton sets. As expected,
the largest differences can be seen at negative $\eta$ values, where
theoretical predictions can vary for about one order of magnitude.
\begin{figure}
\centerline{
\vspace{-0.5cm}
   \epsfig{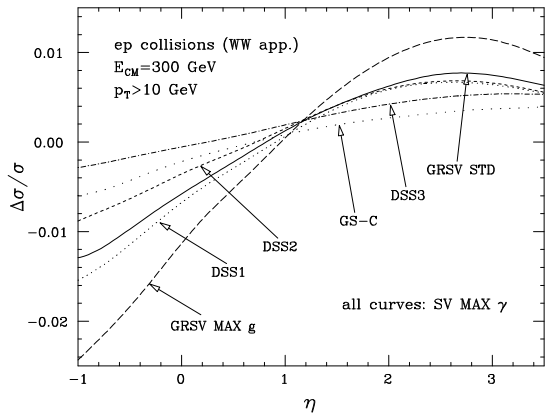} }
\caption{
Same as fig.3 for different polarized proton distributions }
\vspace{-0.4cm}
\end{figure}                                                              
It follows that, if high luminosity will be collected,
it will be possible to get information on the 
polarized parton densities in the proton.  As far as the polarized photon 
densities are concerned, if the ``real'' densities are similar to those 
of the set SV~MIN~$\gamma$, it will be extremely hard to even get the 
experimental evidence of a hadronic contribution to the polarized cross 
section. On the other hand, a set like SV~MAX~$\gamma$ appears to give
measurable cross sections.

In conclusion, we reported the calculation of jet cross-sections
in polarized hadron-hadron and electron--hadron (in the photoproduction regime) collisions, which is accurate to NLO in 
perturbative QCD. For all the observables considered, it has been found
that the 
scale dependence is smaller than that of the LO result.
The inclusion of the NLO terms changes the size of the asymmetries by 20\% in the case of proton-proton collisions at RHIC and, considerably reduces it, in the case of electron-proton collisions at HERA
in the pseudorapidity region where the contamination from the 
hadronic photon contribution is minimal. From our analysis, it is clear 
that the inclusion of the NLO corrections is indispensable in order to 
have reliable quantitative calculations.

Measurements of jet cross-sections with polarized beams at RHIC and HERA will be   fundamental tools in order to extract the polarized gluon distribution in the proton. It is, therefore, worth  emphasizing
 that the theoretical tools for the future NLO analysis of the forthcoming data are already available.
 
It is a pleasure to thank S. Frixione, A. Signer and W. Vogelsang for enjoyable  collaborations.


\begin{thebibliography}{9}

\bibitem{Jets97}  
 S.~Frixione, Nucl. Phys. B507 (1997) 295. 

 \bibitem{sust}
 S.~Frixione, Z.~Kunszt and A.~Signer, Nucl. Phys. B467 (1996) 399. 

\bibitem{jetpp} 
 D.~de Florian, S.~Frixione, A.~Signer and W.~Vogelsang, Nucl. Phys. B539 (1999) 455.

\bibitem{jetep} 
  D.~de Florian and S.~Frixione, to be published in Phys. Lett. B, hep-ph/9904320.


\bibitem{polpar} 
 M.~Gl\"{u}ck, E.~Reya, M.~Stratmann and W.~Vogelsang, Phys. Rev. D53 (1996) 4775;  \\
T.~Gehrmann and W.J.~Stirling, Phys. Rev. D53 (1996) 6100; \\
 D.~de~Florian, O.A.~Sampayo and R.~Sassot, Phys. Rev. D57 (1998) 5803.

\bibitem{svgamma} 
 M.~Stratmann and W.~Vogelsang, Phys. Lett. B383 (1996) 370.


\end{thebibliography}
\end{document}